\documentclass[fleqn,12pt,twoside]{article}
\usepackage{espcrc1}
\usepackage{graphicx}
\usepackage[figuresright]{rotating}

\newcommand{\AmS}{{\protect\the\textfont2
  A\kern-.1667em\lower.5ex\hbox{M}\kern-.125emS}}
 
\title{Evidence for transfer followed by breakup in $^{7}$Li +  $^{65}$Cu}

\author{
A. Shrivastava\address[a]{Nuclear Physics Division, Bhabha Atomic Research Centre, Mumbai 400085, India}
\thanks{email:aradhana@apsara.barc.ernet.in},
A. Navin\addressmark[a]\thanks{Permanent address: GANIL, Bd. Henri Becquerel, BP 55027, Cedex 5, 14076, Caen, France},
N. Keeley\address[b] {DSM/DAPNIA/SPhN, CEA Saclay, F-91191 Gif sur Yvette,  France},
K. Mahata\addressmark[a],
K. Ramachandran\addressmark[a],
V. Nanal\address[c]{DNAP, Tata Institute of Fundamental research, Mumbai 400005, India},
V.V. Parkar\addressmark[a],
A. Chatterjee\addressmark[a]
and S. Kailas\addressmark[a]
}
\begin{document}
 
\maketitle
 \begin{abstract}

The observation of a large  cross-section for the $\alpha + d$ channel
compared to  breakup into the $\alpha  + t$ channel  from an exclusive
measurement for the $^{7}$Li+$^{65}$Cu  system at 25 MeV is presented.
A detailed analysis of the angular distribution using coupled channels
Born approximation  calculations has provided clear  evidence that the
observed  $\alpha +  d$ events  arise from  a two  step  process, i.e.
direct transfer to the 2.186 MeV ($3^+$) resonance in the $\alpha + d$
continuum  of $^6$Li followed  by breakup,  and are  not due  to final
state interaction effects.

PACS: 25.70.Mn, 25.70.Hi, 25.70.Bc, 25.70.Ef, 24.10.Eq

{\it Keywords:  } Exclusive breakup, Transfer-breakup, Elastic scattering,
Weakly bound nuclei, Coupled channels calculations.
\end{abstract}
\newpage
   
\section{Introduction}

 Nuclear  reactions involving unstable/weakly  bound nuclei  that have
 low  breakup  thresholds   and  exotic  structures  display  features
 remarkably  different from  those of  well-bound stable  nuclei.  The
 advent of recent  ISOL facilities, apart from opening  new avenues in
 this field,  has also  caused a  revival in the  study of  stable but
 weakly bound  nuclei like  $^{6,7}$Li and $^9$Be.   Measurements with
 these  nuclei,   with  better  understood   cluster  structures,  are
 relatively easier,  due to the larger available  beam intensities. At
 energies  around the  barrier the  effect  of breakup  on the  fusion
 process and  also the measurement of associated  direct processes are
 topics of current interest.

Recent interpretations  of measurements  of products arising  from the
 excited compound  nucleus (fission or evaporation  residues) show the
 need  for distinguishing  between various  mechanisms leading  to the
 same final product prior to deriving any conclusions about the effect
 of  weak  binding  on  other reaction  processes  \cite{raa04,nav04}.
 Measurements       involving        $^6$He       and       $^{6,7}$Li
 \cite{nav04,dip04,pak03,kol02,kel00},  having  an  alpha  +  {\it  x}
 configuration,  show  significantly  large cross-sections  for  alpha
 particle   production.   It   is  non-trivial   to   disentangle  the
 contributions  to the  alpha  yield arising  from different  reaction
 mechanisms   that  include  fusion-evaporation,   transfer,  transfer
 followed  by evaporation/  breakup of  the ejectile  and  direct (non
 capture)  breakup  of   the  projectile,  especially  from  inclusive
 measurements.  It is a well known observation that the alpha yield is
 much larger than the corresponding triton/deuteron yield in reactions
 involving  $^{6,7}$Li beams  and  some possible  scenarios have  been
 suggested \cite{nav05,sig03,cas78,fle78,uts83}.

 Exclusive  measurements   of  alpha  particles  provide   a  tool  to
 understand the  deconvolution of these  processes. Such measurements,
 along with  differential cross-sections for  various direct processes
 like elastic  scattering and few nucleon  transfer reactions, provide
 important constraints for coupled channels calculations necessary for
 a consistent understanding of reactions with weakly bound nuclei near
 the Coulomb barrier. Apart from  the above stated motivations, one of
 the key  aspects of the present  work is to investigate  the two step
 reaction mechanism, namely, one  nucleon transfer to a resonant state
 followed  by breakup,  for  the  case of  a  $^7$Li projectile.   This
 complex  process needs  the  simultaneous understanding  of both  the
 breakup  and  transfer reactions.   Further,  such measurements  also
 provide  information on  the $^7$Li  wave  function in  terms of  the
 $^6$Li(ground/resonance)$+n$    configuration.     In   an    earlier
 measurement  for  the   $^7$Li+$^{197}$Au  system  \cite{que74},  the
 authors speculated  on the possibility  of such a process  and placed
 limits on the cross-section.  Recently, analysing powers for this type
 of reaction have been reported \cite{nd04}.  In more recent exclusive
 measurements of $\alpha - n  $ for the $^6$He +$^{209}$Bi system, the
 contributions  of single and  two neutron  transfer to  the continuum
 followed by  evaporation have been  reported \cite{kol04,kol05}.  The
 present work  is the  first quantitative measurement  of differential
 cross-section for the transfer-breakup reaction.

 We report  in this letter  exclusive measurements of  alpha particles
along with  $d/t$ to identify  different reaction mechanisms  of alpha
emission, and  a detailed study  of the transfer-breakup  mechanism of
$^7$Li  on  a  medium  mass   target  at  energies  near  the  Coulomb
barrier. These  have been compared  with similar measurements  using a
$^6$Li  beam.  Elastic  scattering angular  distributions  and nucleon
transfer  angular  distributions have  also  been measured.  Extensive
coupled  channels Born  approximation (CCBA)  calculations  along with
continuum   discretized  coupled   channels  calculations   (CDCC)  to
simultaneously explain the large number of observables are presented.
 
\section{Experimental details and results}
The  measurements  were performed  at  the  14UD BARC-TIFR  Pelletron,
Mumbai using 25 MeV $^{6,7}$Li  beams on an enriched $^{65}$Cu target.
The  target thickness used  for measuring  the elastic  scattering and
nucleon-transfer differential cross-sections  was 1.0 mg/cm$^2$, while
for the exclusive breakup it  was 2.5 mg/cm$^2$. Typical beam currents
were  around 2  pnA.  The  coincidence detection  system  consisted of
three  $\Delta$E(30,  40 and  47$\mu$m  )--E(2mm)  Si surface  barrier
telescopes  and a  $10\times 10\times 10$mm$^3$ CsI(Tl)  charged  particle detector,
kept 20$^{\circ}$  apart.  With the  present setup it was  possible to
measure fragment  kinetic energies corresponding to  unbound states of
$^{6,7}$Li up to  an excitation energy of 5.5  MeV.  From the measured
kinetic  energy  of each  fragment  and  the  breakup $Q$  value,  the
relative  energies between  $\alpha-d(t)$  for $^6$Li($^7$Li)  breakup
were deduced  using three body final  state correlations \cite{sch77}.
The detectors were calibrated using discrete energy alpha particles in
the range of 12 to 26  MeV \cite{ser73}, produced in the reaction with
$^7$Li beam  of energies  20 and  25 MeV on  a 50  $\mu$g/cm$^2$ thick
$^{12}$C target.  The elastic  scattering, transfer and alpha emission
angular  distribution  measurements   were  performed  in  a  separate
experiment employing three  Si-surface barrier telescopes ($\Delta$E--
10,  20 and  25  $\mu$m and  E--  1mm) covering  an  angular range  of
10$^{\circ}$  to 140$^{\circ}$.  A  300 $\mu$m  thick Si-  detector at
forward angles, for both the exclusive and inclusive measurements, was
used for monitoring the number of incident beam particles.

The data  were collected in an  event by event mode,  with the trigger
generated from fast coincidences  between adjacent detectors.  In Fig.
1  alpha and  deuteron coincidence  spectra  for both  the systems  at
$\theta_{\alpha  d}$  = 20$^{\circ}$  are  shown.   The two  localized
contributions  in   the  spectra  are  identified   using  three  body
kinematics \cite{sch77},  and correspond to the  sequential breakup of
the first resonant  state in $^6$Li.  The full curves  in Figs. 1a and
1b show the  calculated kinematic correlations for the  breakup of the
3$^+$ state in  $^6$Li.  In Fig. 1b, the  relative energy ( E$_{\alpha
d}$) of 0.71 corresponding to  the 3$^+$ resonance state is indicated.
Fig.  1c shows the projection of Fig.  1b along the alpha energy axis.
The alpha particles moving forward (backward) in the $\alpha-d$ center
of  mass system  corresponds  to the  high  (low) energy  peak in  the
spectrum.

Shown  in Fig. 2a  are the  angular distributions  for the  breakup of
$^6$Li proceeding through the 3$^+$ and 2$^+$ resonant states at 2.186
and  4.312   MeV,  respectively.   Differential   cross-sections  were
computed from  $\alpha$ +  $d$ coincidence yields,  assuming isotropic
emission of breakup fragments in the $^6$Li$^*$ frame and are shown in
Fig.  2.  The Jacobian of  transformation and the center of mass angle
of the  scattered $^6$Li$^*$ were obtained as  described in references
\cite{mei85,fuc82}.  The ground state and first resonant state (3$^+$,
2.186  MeV)  of  $^6$Li  are  well  described  with  a  $\alpha  +  d$
configuration \cite{til02}.   The decay  of the  second  resonant state
(0$^+$, 3.56 MeV) of $^6$Li to $\alpha + d$ is forbidden due to parity
considerations; however, a  decay through the $t$ +  $^3$He channel is
possible.   The  cross-section  for  this  state was  inferred  to  be
negligible as $^3$He + $t$ coincidences were not observed.

The importance of the two step  process can be seen from Fig. 1d which
shows the $\alpha $+  $d$ coincidences for the $^7$Li+$^{65}$Cu system
at 25  MeV. Two clear peaks are  seen in the deuteron  vs alpha energy
correlation  plot,  indicating  breakup  of $^6$Li  formed  after  one
neutron stripping ($Q$ = -0.185 MeV) of $^7$Li and clearly not breakup
of  $^7$Li into  $\alpha+d+n$ ($Q$  =  -8.8 MeV).   From the  relative
energy plot  shown in  Fig.  1e,  the two peaks  can be  identified as
arising  from the breakup  of $^6$Li  via its  2.18 MeV  state, formed
after  a one neutron  transfer reaction.   The transfered  neutron can
populate  various states  in  $^{66}$Cu depending  on  their spin  and
spectroscopic factors.   The kinematic curves shown in  the figure are
for  transfer of  the neutron  to the  ground, 1.15  MeV and  2.14 MeV
states of $^{66}$Cu  with $^6$Li in its 3$^+$  resonance state. As can
be seen from the figure, there is very good agreement of the data with
these  kinematic  plots.    The  corresponding  angular  distribution,
integrated over  excited states  of $^{66}$Cu ,  is displayed  in Fig.
2b.  Also  shown in the figure  are the angular  distributions for the
breakup  of $^7$Li  $\rightarrow$  $\alpha$ +  $t$  through its  first
resonance  state at  4.63 MeV.   From the  known cluster  structure of
$^7$Li one would have naively expected a much larger cross-section for
the latter compared to the $\alpha$ + $d$ coincidence yield.

 The measured elastic angular distributions for the $^6$Li + $^{65}$Cu
and  $^7$Li  +  $^{65}$Cu systems  are  shown  in  Figs.  2c  and  2d,
respectively.  The errors  on the data points shown  in the figure are
statistical only.   The  angular  distribution for  the  one  neutron
stripping of $^7$Li + $^{65}$Cu  (Q = -0.185 keV) populating $^6$Li in
its ground state (as $^6$Li has no bound excited states) and $^{66}$Cu
(E$^*$ up  to 5 MeV)  was obtained independently from the inclusive data.
The integrated  cross-section obtained  assuming a Gaussian  shape for
the angular  distribution is listed in  Table 1.  As can  be seen from
Table  1, this cross-section  is larger  than that  for all  the other
direct processes  listed.  The  errors on the  measured cross-sections
are  from  uncertainties in  the  fitting  procedure  (6 to  8\%)  and
statistics   (3   to   6\%).    The  1$n$-pickup   and   $t$-stripping
cross-section for $^6$Li + $^{65}$Cu  are also listed in Table 1.  For
both isotopes the exclusive breakup cross-section  is observed to
be   much  smaller   than  the   inclusive  cross-section   for  alpha
emission. The contribution  of alpha particles evaporated from  the compound nucleus
is estimated  to be less  than 30\% (Table  1) of the  total inclusive
alpha  yield.  These were  obtained by  fitting the  measured backward
angle alpha angular distribution using the statistical model code PACE
\cite{gav80}.  This  suggests that the majority of  the alpha particle
yields arise from processes  where the deuteron (triton) is transfered
or captured  by the target  \cite{nav05,sig03,cas78,fle78,uts83} after
breakup  of the  $^6$Li($^7$Li) in  field  of the  target.  The  alpha
emission cross-section from the inclusive and exclusive data
 for the $^6$Li projectile is  larger than that
for $^7$Li, as expected from the difference in the breakup thresholds.
Similar results  for the inclusive alpha emission cross-sections  for $^6$Li and
$^7$Li were obtained in references \cite{kel00,pfe73}.

\section{Calculations}\label{sec:calcs}

Two distinct sets of calculations were carried out to describe the data.
Those for  the elastic scattering  and breakup
processes were performed within the CDCC formalism using a cluster
folding  model  \cite{buc77} for  the  structure  of $^{6,7}$Li. Calculations
for the transfer  breakup employed the  CCBA framework,
i.e.  CDCC in the entrance and  exit channel and DWBA for the transfer
step, utilizing  the potentials that explained  the elastic scattering
data.  Both  the CCBA and  CDCC calculations described here  were performed
using the code FRESCO \cite{ijt}.

The CDCC calculations for  $^6$Li ($^7$Li) were similar  to those
described  in \cite{kee03,kee02} and assumed an $\alpha$+$d(t)$  cluster
structure.  The  binding potentials between  $\alpha$+$d$ and
$\alpha$+$t$  were taken from  \cite{kh} and  \cite{bm}, respectively.
The  $\alpha$+$d(t)$  continuum  was  discretized  into  a  series  of
momentum bins  of width $\Delta k =  0.2~ $fm$^{-1}$ (up to  $k$ = 0.8
fm$^{-1}$) for  $L$ =  0, 1, 2  for $^6$Li and  $L$ =  0, 1, 2,  3 for
$^7$Li , where $\hbar$$k$ denotes  the momentum of the $\alpha + d(t)$
relative   motion.   All  couplings,   including  continuum--continuum
couplings,  up  to  multipolarity  $\lambda =  2$  were  incorporated.
Optical  potentials for  $\alpha$+$^{65}$Cu  and $d(t)$+$^{65}$Cu  are
required  as  input  for the  cluster-folded  $^6$Li($^7$Li)+$^{65}$Cu
potential.  The  potential  for  $\alpha$+$^{65}$Cu  was  obtained  by
adjusting the real and imaginary depths of the global $\alpha$ optical
potential of Avrigeanu {\em et al.\/} \cite{av}, in order to match the
measured  $\alpha$+$^{65}$Cu  elastic  scattering  data  of  reference
\cite{cos}.  In  the absence of suitable elastic  scattering data, the
global parameters of \cite{bg,pp}  were employed for the $t$+$^{65}$Cu
and  $d$+$^{65}$Cu   optical  potentials.   The   real  and  imaginary
strengths of  the cluster-folded potentials were  adjusted to describe
the  elastic scattering  data.  

Results  of the  calculations  for the
elastic scattering are shown in Figs. 2c and 2d.  The two different curves
are  results  of the  calculation  performed  with  (solid lines)  and
without   (dashed   lines)    couplings.    The   calculated   angular
distributions  for the  resonant  states, 3$^+$  and  2$^+$ of  $^6$Li
(Fig. 2a)  and 7/2$^-$ of $^7$Li  (Fig.  2b) show  good agreement with
the data. The results of the full CDCC calculation are listed in Table
1.  The  angle  integrated   cross-sections  of  the  resonant  states
obtained  by fitting to  a Gaussian  distribution show  good agreement
with the calculation. The  total calculated breakup cross-sections for
$^6$Li and $^7$Li were  obtained by integrating contributions from the
states in  the continuum up to 11 MeV. As can be  seen in Table  1, the
total $^7$Li($^6$Li) $\rightarrow$  $\alpha$ + $t(d)$ breakup provides
a negligible contribution to the total reaction cross-section.
The fusion cross-sections listed in Table 1, for $^{6,7}$Li +  $^{65}$Cu
were obtained using the barrier penetration model, as described in
reference \cite{nick02}. 

As mentioned earlier, for the  $^7$Li + $^{65}$Cu system a significant
number of $\alpha$+$d$ coincidence events consistent with decay of the
2.18 MeV $^6$Li($3^+$) resonance  were observed. The simplest reaction
mechanism for  producing these events is  a transfer--breakup process,
with direct  neutron stripping to  the unbound $3^+$  resonance and/or
neutron  stripping  to  the  $^6$Li  $1^+$ ground  state  followed  by
excitation  to the  $3^+$ resonance  through final  state interaction.
Due to  the high density of  states in the  residual $^{66}$Cu nucleus
and   the  experimental  resolution,   the  angular   distribution  of
$^6$Li($3^+$)  resonance  events  was  integrated  over  the  residual
$^{66}$Cu excitation energy up to 2.5 MeV. The $^{66}$Cu could thus be
left in any  one of up to 40 states \cite{dp69}.   

It was not possible
to incorporate  this process directly into the  full CDCC calculation.
Thus, to establish  the dominant reaction mechanism  for the
observed  $\alpha$+$d$  coincidences  in  $^7$Li +  $^{65}$Cu  (direct
transfer  to the  continuum or  transfer to  ground state  followed by
breakup),  a series  of  CCBA calculations  employing  a much  reduced
coupling scheme in the entrance  and exit partitions, shown in Fig.  3
was performed. The potentials used were taken from the CDCC
calculation as  explained in the previous paragraph.   For the $^6$Li+
$^{66}$Cu exit  partition, only coupling  to the  2.18 MeV  ($3^+$) of
$^6$Li was retained.  The optical potential for $\alpha$+$^{66}$Cu was
again  calculated from  the  global parameters  of  Avrigeanu {\em  et
al.\/} \cite{av}, renormalized  by the same factors needed  to fit the
14  MeV $\alpha$+$^{66}$Cu data  of Costa  {\em et  al.\/} \cite{cos}.
The optical  potential for $d$+$^{66}$Cu  was the central part  of the
potential of  Bieszk and Knutson  \cite{bie} for 9  MeV $d$+$^{63}$Cu.
Spectroscopic factors for the $^7$Li $\rightarrow$ $^6$Li+$n$ overlaps
were taken from Cohen and  Kurath \cite{ck}.  The neutron was bound in
a  Woods--Saxon  well  of   radius  1.25  $\times$  A$^{1/3}$  fm  and
diffuseness 0.65  fm, the  depth being adjusted  to yield  the correct
binding energy.  The spectroscopic factors for $^{66}$Cu $\rightarrow$
$^{65}$Cu+$n$  were  taken from  Daehnick  and  Park \cite{dp69}.  The
neutron was again bound in a Woods--Saxon well of radius 1.25 $\times$
A$^{1/3}$ fm  and diffuseness 0.65  fm, as used in  ref.\ \cite{dp69}.
The transfer part of the calculations was performed using the post--form DWBA and
included the full complex remnant term.

The  CCBA  calculations were  carried  out  for  transfers leaving  the
residual $^{66}$Cu in  levels up to 1.43 MeV in excitation, partly due
to the uncertain nature of many  of the spin assignments above 1.5 MeV
and the presence  of unresolved doublets. As the  reaction Q-value for
$^{65}$Cu($^7$Li,$^6$Li)$^{66}$Cu is slightly  negative (-0.185 MeV), 
the population of states near the ground state of $^{66}$Cu will be
favoured to some extent due to Q-matching considerations.  Hence  the sum of the
present CCBA calculations covers  most of  the observed  $\alpha$+$d$
coincidence cross-section.  In the cases where the spin assignments of
Daehnick and  Park differ from  those of the  compilation \cite{bhat},
the latter has been followed.  The  shape of the calculated sum of the
CCBA angular  distributions is in good agreement  with the measurement
(Fig.  2b),  although the  magnitude is lower  due to the omission  of
$^{66}$Cu  states above  1.43  MeV.  The  results  of the  calculation
confirm the  transfer/breakup mechanism for  the observed $\alpha$+$d$
coincidences.  Normalising  the summed  CCBA calculations to  the data
yields a  total cross-section for $^6$Li($3^+$) production  of about 9
mb, nearly twice  that for the measured breakup  of $^7$Li via 7/2$^-$
state. The seperately measured $^{65}$Cu($^7$Li,$^6$Li)$^{66}$Cu reaction
leaving $^{6}$Li in its  ground state is also well described. The dominant
peak in the spectrum for this transfer is centred on the 1.15 MeV $6^-$ state
in $^{66}$Cu, and the experimental value for the total cross section
integrated over a bin of width 400 keV centred at 1.15 MeV is $4.2 \pm 0.5$
mb, while the summed total cross section from the CCBA calculation for
$^{66}$Cu states in the same energy range is 3.9 mb. 

Having established  that the  $^6$Li($3^+$) resonance is  populated by
the transfer--breakup mechanism,  a further distinction between direct
transfer to  the unbound  $3^+$ resonance in  $^6$Li (transfer  to the
continuum) and transfer  to $^6$Li in its $1^+$  ground state followed
by excitation to the $3^+$ (final state interaction) was investigated.
Calculations omitting the  direct transfer step  showed that
the final state interaction process provides a negligible contribution
(10\%) except  at extreme  forward angles.  It  can thus  be concluded
that the main reaction mechanism for the observed large $\alpha$ + $d$
exclusive cross-sections is direct transfer followed by breakup of the
unbound $3^+$ resonant state in the $^6$Li continuum.

\section{Discussion}

 The  exclusive  breakup cross-sections  for  the  resonant states  of
 $^{6,7}$Li  could be  explained well  by CDCC  calculations performed
 using   potentials   that   fit   the  elastic   scattering   angular
 distributions.   The  total  non  capture breakup  cross-section  for
 $^6$Li was  found to  be larger  than that for  $^7$Li mainly  due to
 the lower  alpha binding  energy in
 $^6$Li compared to $^7$Li.   The exclusive breakup cross-sections are
 a very small fraction of  the reaction cross-sections for both $^6$Li
 and  $^7$Li  (Table  1).   The  exclusive breakup  for  $^{6,7}$Li  +
 $^{65}$Cu contributes less than  10\% and compound nucleus evaporation
 less   than   30\%   towards   the   observed   large   alpha-singles
 cross-section. The origin of  the large alpha yield in $^6$Li($^7$Li)
 induced  reactions  seems  to   be  mainly  due  to  deuteron(triton)
 capture/deuteron(triton)      transfer      as      discussed      in
 \cite{nav05,sig03,cas78,fle78,uts83}.

In a recent  study with $^{6,7}$Li on a  medium mass target, $^{64}$Zn,
 very large cross-sections for the  break up (where both the fragments
 survive) have  been indirectly  inferred by subtracting  the complete
 and incomplete fusion  cross-sections from the reaction cross-section
 \cite{gom04}.   This  could arise  from  a  neglect  of other  direct
 reaction   processes,  for   instance;  nucleon   transfer,  inelastic
 excitation  of the  target/projectile  etc.  Before  arriving at  any
 conclusion  on  the  role  of  breakup  on  other  reaction  channels
 unambiguous  information on the  breakup cross-section  is necessary.
 The present  work clearly shows  that exclusive measurements  for the
 breakup  cross-sections are  essential  and indirect  methods can  be
 unreliable.

\section{ Conclusion}

  The present work reports a  detailed study of the multi-step reaction
mechanism, namely transfer-breakup. The origin of the large yields for
$\alpha + d$  events from the coincidence data  for $^7$Li breakup has
been identified as transfer followed  by breakup of the excited $^6$Li
via  its 3$^+$  resonant  state in  the  continuum.  To  get a  deeper
insight into the mechanism behind  this reaction -- direct transfer of
the neutron to  the $^6$Li--continuum or transfer to  the ground state
of $^6$Li followed by excitation to the continuum -- CCBA calculations
were performed.  The results of the calculations have established that
the main reaction mechanism is direct transfer to the continuum.

   Reactions with low energy unstable radioactive ion beams from newly
available  facilities  are  expected  to  be  of  similar  complexity.
Identification   of  the   reaction  processes   and   development  of
theoretical   understanding  for   such  multi-step   reactions   is  a
challenging task.   The present study with weakly  bound stable nuclei
on  breakup  and  transfer--breakup  mechanism  along  with  extensive
theoretical analysis is an attempt in this direction.

\begin{table}[htb]
\begin{center}
\caption{ Cross-sections for various channels in  $^6$Li + $^{65}$Cu and $^7$Li + $^{65}$Cu systems.
 The calculated values  are result of coupled channel calculations (see text).}
\vspace {1cm}
\begin{tabular}{@{}lllllll}
 $^6$Li + $^{65}$Cu& &\\
\hline
Channel  & $\sigma_{exp}$ (mb) & $\sigma_{cal}$ (mb)  \\
\hline
$^6$Li$^*$(2.186 MeV)$\rightarrow$ $\alpha+d$ & 22 $\pm$2 &  19.5  \\
$^6$Li$^*$(4.31 MeV)$\rightarrow$ $\alpha+d$ & 4.3 $\pm$ 0.5 &  3.9 \\ 
$^6$Li$^*$(5.65 MeV)$\rightarrow$ $\alpha+d$  & - & 0.8                       \\
$^6$Li$^*$(upto 11 MeV) $\rightarrow$ $\alpha+d$   & -  &48 \\
 $^7$Li (1-neutron pickup)       & 14.7 $\pm$ 2.0 & -  \\
 $^3$He (triton stripping)      & 3.3  $\pm$ 0.5 & -  \\
 $\alpha$ (CN evaporation) & 177 $\pm$ 20 & -   \\
 $\alpha$ (inclusive) & 612 $\pm$ 40 & -   \\
Fusion                &  - & 1199            \\
Total reaction        &  - & 1492            \\

\hline\\
 $^7$Li + $^{65}$Cu& & \\
\hline
 Channel & $\sigma_{exp}$ (mb) &  $\sigma_{cal}$ (mb) \\
\hline
 $^7$Li$^*$(4.652 MeV)$\rightarrow$ $\alpha+t$ & 4.5 $\pm$ 0.6 &  5.1 \\
 $^7$Li$^*$(7.454 MeV)$\rightarrow$ $\alpha+t$ & - & 0.4 \\ 
 $^7$Li$^*$(upto 11 MeV)$\rightarrow$ $\alpha+t$ & - & 20.9\\
 $^6$Li$^*$(2.186 MeV)$\rightarrow$ $\alpha+d$ & 9 $\pm$ 1 &5.6 \\
 $^6$Li (1-neutron stripping) & 44 $\pm$ 4 & 9.3\\
 $^6$He (1-proton stripping)  & 7.8 $\pm$ 1.0 & -\\
 $\alpha $ (CN evaporation) & 110 $\pm$ 18 & -   \\
 $\alpha$ (inclusive) & 422 $\pm$ 33 & -  \\
Fusion                &  - & 1061            \\
Total reaction      &  - & 1401           \\

\hline
\end{tabular}
\end{center}
\end{table}
\begin{figure}[ht]
\begin{center}
\includegraphics[width=30pc]{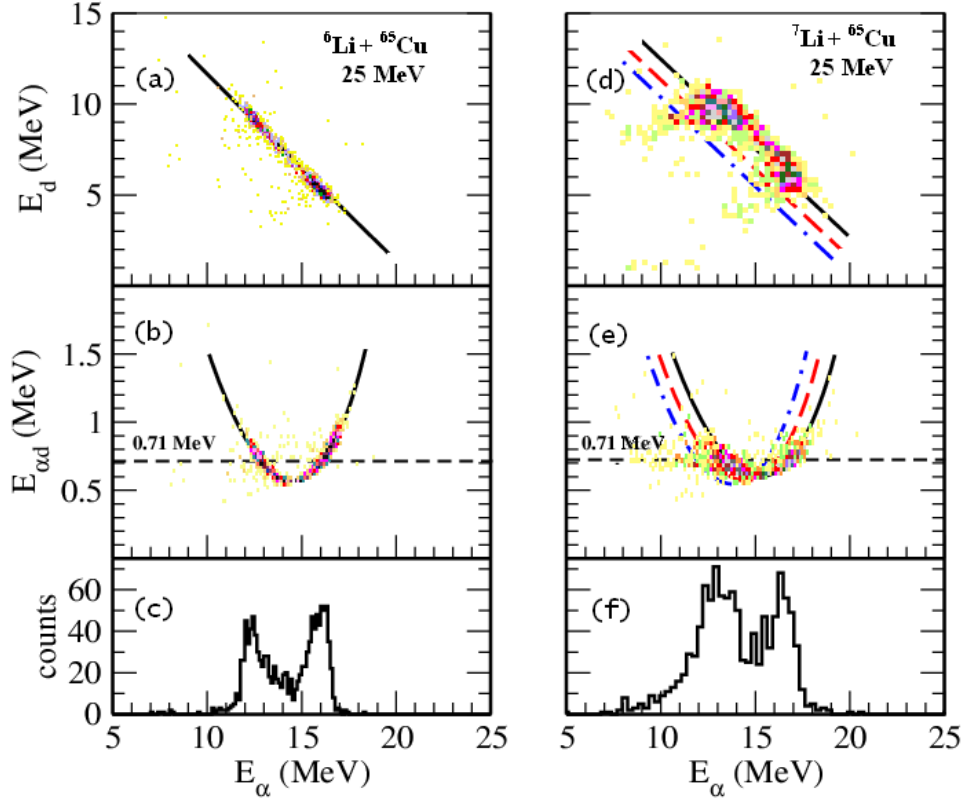} 
\it\caption{\label{fig1}(Color online) Alpha
- deuteron correlations  for $^{6,7}$Li  + $^{65}$Cu systems.   For the
$^6$Li  projectile the  data are  for $\alpha$  particles  detected at
65$^{\circ}$ and deuteron at 45$^{\circ}$ plotted as (a)E$_d$ vs E$_{\alpha}$
(b) the relative  energy E$_{\alpha d}$ vs  E$_{\alpha}$ and
(c) projection of the $\alpha$  particle energy for data shown in (b).
The solid curves in (a) and  (b) are results of three body kinematical
calculations.  Similar plots for  $^7$Li projectile for the breakup of
$^6$Li$^*$ after a 1n stripping reaction are shown in (d), (e) and (f)
for  $\alpha$  particles  detected  at 26$^{\circ}$  and  deuteron  at
46$^{\circ}$. The solid,  dashed and dot-dashed curves in  (d) and (e)
are the  same as  above corresponding to  $^{66}$Cu in the  ground and
excited states at 1.15 and 2.14 MeV respectively.}
\end{center}
\end{figure}

\begin{figure}[ht]
\begin{center}
\includegraphics[width=30pc]{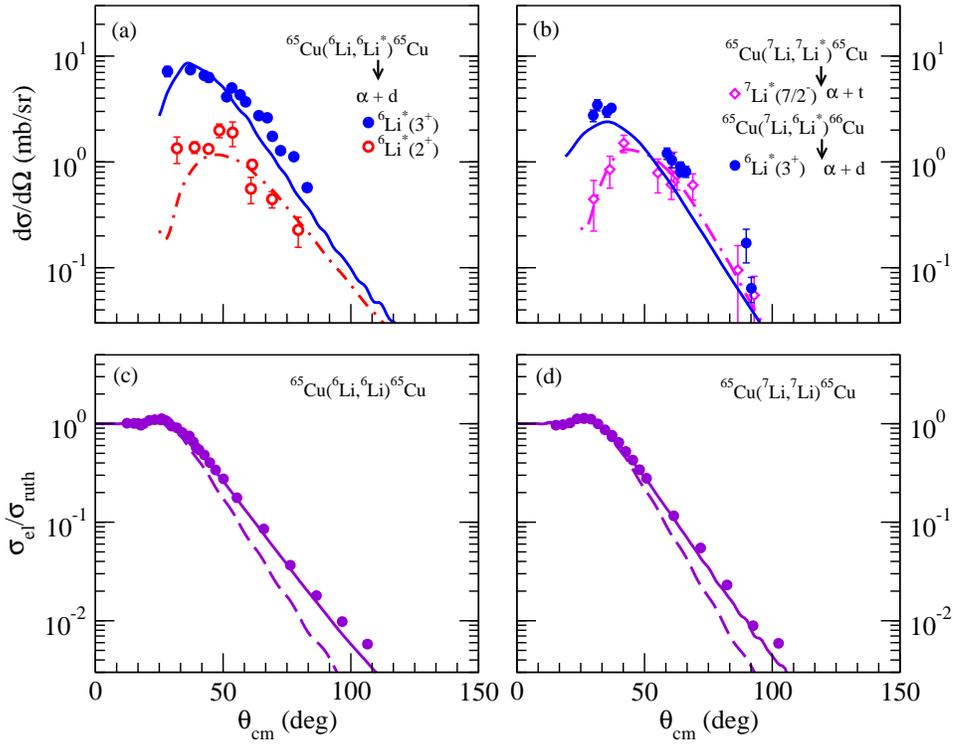}
                                                                                
\it\caption{\label{fig2}  Differential  cross-sections  for  resonant,
transfer  breakup and  elastic scattering.   (a) Breakup  via resonant
states  2.18  MeV  (3$^+$)  and  4.31  MeV  (2$^+$)  in  $^6$Li  (CDCC
calculations are shown as solid  and dash-dot lines).  (b) Breakup via
resonant  state   4.63  (7/2$^-$)  MeV  in  $^7$Li   along  with  CDCC
calculations (dash-dot lines) and data for transfer--breakup reaction,
$^7$Li +  $^{65}$Cu$\rightarrow$ $^6$Li$^*$(3$^+$) +  $^{66}$Cu$^*$ (0
to 2.5  MeV) along with CCBA  calculations.  (c) and (d)  The ratio of
the elastic  scattering to the Rutherford cross-section  as a function
of  angle  for  $^6$Li  +  $^{65}$Cu and  $^7$Li  +  $^{65}$Cu.   CDCC
calculations  are  shown as  solid  (coupled)  and dashed  (uncoupled)
lines.}
\end{center}
\end{figure}

\begin{figure}[ht]
\begin{center}
\includegraphics[width=30pc]{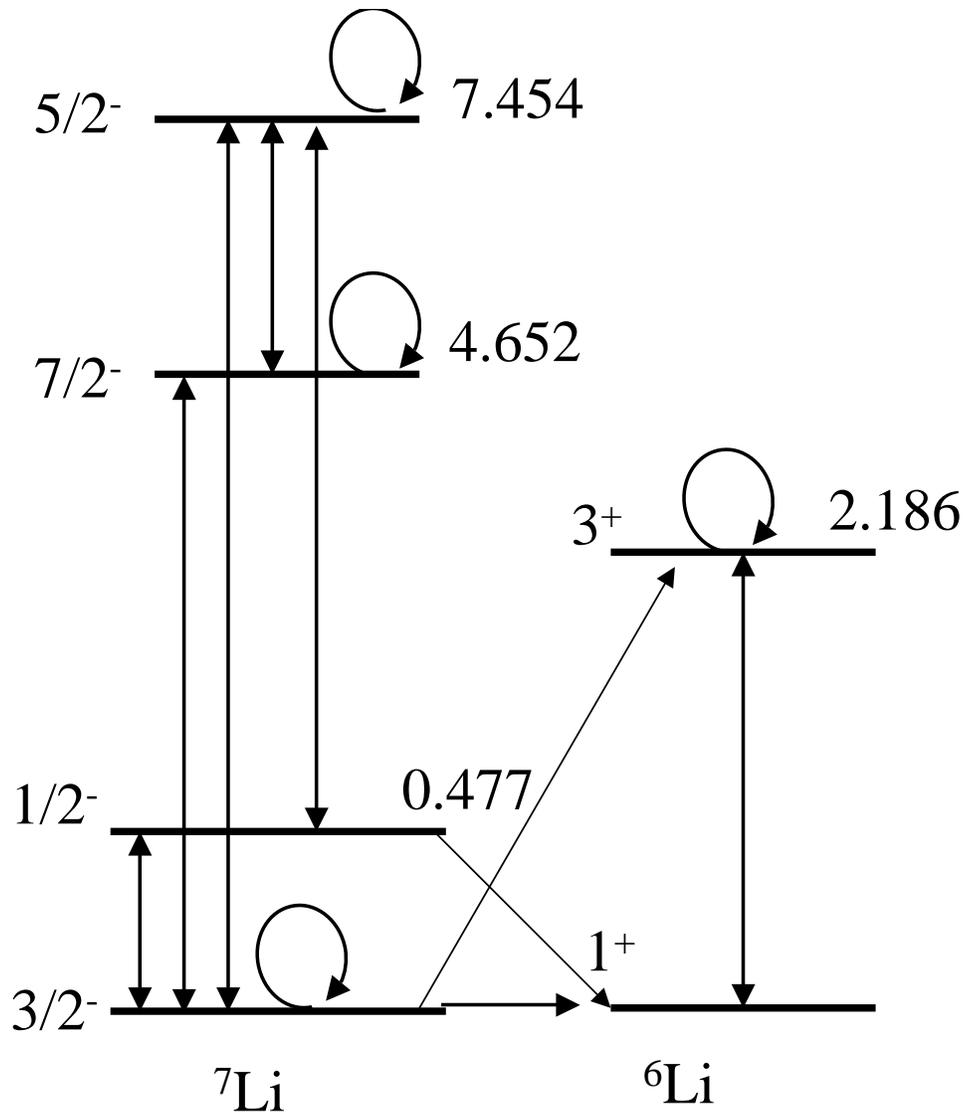}
                                                                                
\it\caption{\label{fig3} Reduced coupling scheme for the projectile
used in  the CCBA calculations (see text).}
\end{center}
\end{figure}

\end{document}